*Opinion*

# Choices in the transformative Anthropocene

**Miguel Pinheiro[1] & Pablo Pena Rodrigues[1*]**

[1]Jardim Botânico do Rio de Janeiro, Rua Pacheco Leão, 915, CEP 22460-030, Rio de Janeiro, Brazil

*Corresponding author: Rodrigues, P. P. (pablojfpr@hotmail.com )

**Abstract**

Unprecedented imbalances and growing human impacts characterize the Anthropocene. It highlights the urgency of better choices, and the perspectives outlined here can inform our decision-making process. The *"Biocentric-Technological"* way stresses the need to change the human niche to recalibrate human-ecological interactions in order to halt the process of biosphere transformation. The *"Bio-Anthropogenic"* way is a middle path between the extremes, based on strategies that could lead to novel ecosystems and a symbiotic relationship between humans and new organisms. This will require ethical, cultural and technological changes towards a less harmful transformation. The final is the *"Anthropocentric"* way, which is the outcome of current human behavior and population growth. Failure to develop new environmentally friendly technologies and cultures will lead to the collapse of current life-support systems, leading to hazardous scenarios. These perspectives can help us to choose a safer and more sustainable future.

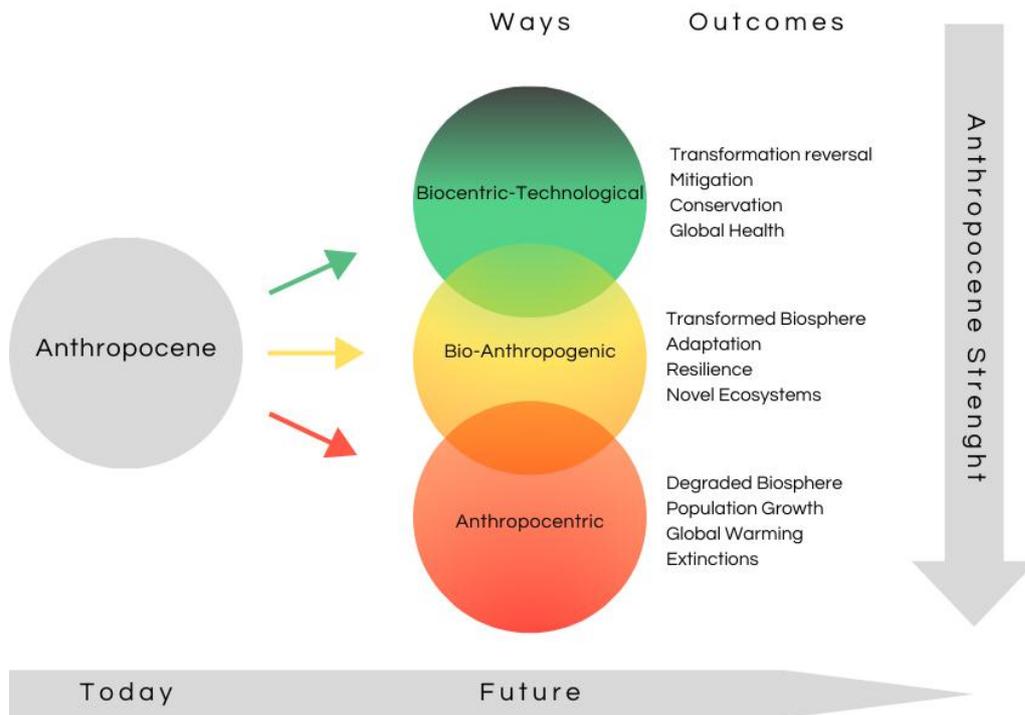

Three potential pathways ("Ways") and their outcomes shaped by the strength of the Anthropocene over time.

## Introduction

There is strong evidence that human activities are changing the biosphere, and recent advances have made it possible to detect and quantify these processes (Rockström et al. 2009; Steffen et al. 2015). Once the changes were recognized, there was a shift towards seeking solutions (Wakefield 2017; Jeanson et al. 2020; Abhilash 2021; McPhearson et al. 2021). The current crisis requires us to rethink the way we live (Ly and Cope 2023). Despite the efforts of many individuals and groups to find innovative solutions (Scheffran et al. 2024), the changes needed to achieve a better future face resistance. Wars, environmental disasters, pandemics, climate change and other impacts do not seem to be enough to change human behavior (Allen et al. 2024). Conversely, the dominant economic and social systems seem to benefit from these crises. Many advances have their roots in war technologies, which are primarily designed to perpetuate structures of power and domination (Chin 2019). These patterns of exploitation are a common feature of human history and have now reached sophisticated forms. Some of these technologies also paradoxically contribute to human well-being by facilitating the development of other technologies related to human health and food production (McClements et al. 2021). And so the human population is still growing, even in the face of so many challenges and inequalities.

The biosphere is the arena in which these processes occur, and the result is profound transformation (Ellis et al. 2021; Chure et al. 2022). It is clear that its ecosystems have capacity to adapt, given their inherent dynamism and susceptibility to variations over time and space. These changes occur as a result of short, medium and long-term environmental changes (Kimmins 2004), which may have anthropogenic causes (Rosenzweig et al. 2008). These changes create new environmental dynamics that provide opportunities for biological processes to reorganize. Human activity is a key driver of these changes (Albuquerque et al. 2018; Illy and Vineis 2024), as evidenced by extinctions, anthropogenic processes in industrial economies, population growth and the expansion of transport networks (Boivin et al. 2016).The human species is changing the course of evolution, affecting other species. It will also act as an intense modulator of natural and artificial selection (Sullivan et al. 2017). One of the main consequences of human expansion is environmental degradation. It is therefore imperative to restructure the conceptual framework of human behavior to reduce environmental degradation and climate change. It seems clear that a "viable world" (*sensu* Scheffran et al. 2024) cannot lead to solutions that keep humanity's current anthropocentric way of life.

 **Human niche and environmental changes**

To understand the expansion of the human niche and the way it shapes environments and ecosystems (Odling-Smee et al. 2013), it is necessary to know human nature in depth. Moreover, our niche is based not only on the evolutionary strategy of cognition and genetic inheritance, but also on the ability to develop and shape culture (Fuentes 2017). Our evolutionary trajectory has been linked to the increase in brain mass and developing complex societies (Van Schaik 2016). Knowledge, the use of extra-somatic materials, and resilience are some of the greatest distinctions of humans (Fuentes 2016), which alter their own evolutionary trajectory and create scenarios that lead to adaptive change (Sullivan et al. 2017), acting as a 'hyper-keystone species' (Worm and Paine 2016). In short, humans are niche creators (Albuquerque et al. 2018) and are causing major changes in the biosphere, including: extinctions, extirpations, and changes in species composition, in addition to changes in community structure and diversity. However, environmental functions can be restored through less destructive lifestyles. Current human ways of managing planetary resources are largely associated with consumerism and unnecessary waste (Sehrawat et al. 2015; Zalasiewicz et al. 2017), and it is fundamental to redesign these perspectives.

**Ecological Restoration**

Restoration aims to return ecosystems, and species, to their original state. This is difficult to do because conditions and information are often not available. To make restoration feasible, ecosystems need to be maintained and balanced (Arroyo-Rodríguez et al. 2015; Abhilash 2021). This promotes biodiversity and contributes to ecological restoration (Costanza et al. 1997). Restoration strategies vary (Barral et al. 2015) and natural regeneration can be effective (Uriarte and Chazdon 2016). However, many factors are involved and human intervention is often required (Meli et al. 2017). Passive restoration allows nature to recover in a more spontaneous way (Shono et al. 2007). Active restoration involves planting seedlings and changing logging and burning practices. Strong human intervention can have important consequences for restored communities, such as the establishment of new tree communities. By understanding the dynamics of degraded areas, we can outline a restoration strategy that takes into desired outcomes and remnant vegetation (Chazdon 2008). Restoration can therefore be an excellent alternative for mitigating and even reversing the harmful effects of human activities.

**What choices do we have?**

Through these ideas we can see how our choices will influence outcomes in the future. It is true that human environmental impact is dynamic, so different lifestyle trends need to be considered, as well as how human knowledge will evolve. According to these choices, we have identified 3 main ways: the *"Biocentric-technological"*, the *"Bio-Anthropogenic"* and the *"Anthropocentric"* (Fig. 1). These paths involve technological, cultural and ethical improvements and can help us to choose a path of hope and prosperity.

*The Biocentric-Technological way*

This approach would reorient the human niche towards new ecological relationships. Friendly and respectful interactions with other species can reinvent our way of life. It aims to balance ecosystems and to develop and use new technologies to reduce anthropogenic impacts (Meroni 2007). The potential for knowledge assimilation represents our differential, because through cultural and ethical shifts it would be possible to synthesize significant changes in human behavior (Fuentes 2016). This will require new models of human settlement and agriculture. Human population growth and birth control are also important issues (Speidel et al. 2007).

In terms of ecological restoration, resources would be directed towards promoting active strategies for global regeneration (Holl and Aide 2011) and restoring interspecific interactions (Hobbs et al. 2011), especially in areas of low resilience. Furthermore, agroforestry models are necessary for the advancement of the socio-ecological system, as they would allow the economic transfer of resources while supporting the maintenance of ecosystems (Noordwijk et al. 2020). Another way to protect the environment is to prioritize local food production, in order to reduce the need to buy products from agribusiness, which are harmful to the environment (Steinfeld and Gerber 2010; Berti and Mulligan 2016). Finally, there is a need to rethink the knowledge and technologies aimed at mitigating anthropogenic impacts (Lema and Lema 2012), especially in relation to resource management (Sehrawat et al. 2015) and pollutant emissions. Large-scale deployment of green technologies (Krass et al. 2013) would significantly reduce emissions of pollutants and waste.

*The Bio-Anthropogenic way*

The hallmark of humans in this approach is its influence as a powerful ecosystem engineer, acting as a "hyper-keystone species" (Worm and Paine 2016). Ecological modulation builds new evolutionary pathways and new organisms (Sullivan et al. 2017), creating unprecedented ecosystem dynamics and resilience for certain species (Hobbs et al. 2009), which can be considered anthropogenic-favored. The passive restoration can be used, as the spontaneous recovery would establish a new community dynamic (Shono et al. 2007), creating new emergent ecosystems (Hobbs et al. 2006) adapted to these new conditions. The increase in resilience would occur by incorporating the new interspecific relationships (Elmqvist et al. 2003). In this way, exotic or invasive species are not seen only as negative for ecosystems. Then, in extremely anthropogenic ecosystems, such species may be present as survivors favored by human activities (Pena-Rodrigues and Lira 2019). Moreover, hybrids would represent an important form of variation for the speciation process and the emergence of evolutionary lineages in scenarios of human influence (Forabosco et al. 2013; Barrera-Guzmán et al. 2017), once these organisms have a high evolutionary capacity and evolve rapidly in isolation conditions (Schumer et al. 2015). Indeed, many hybrids with a history of establishment in natural environments have a higher adaptive capacity than their parental species (Crispo et al. 2011).

The emergence of genetically modified organisms is highlighted, both to ensure food security (Zhang et al. 2016) and as a functional adaptation to new scenarios (Catarino et al. 2015). The spread of GMOs would provide the opportunity for the emergence of organisms genetically adapted to new situations created by humans. This is favored by advances in modern biotechnology, not only in plantations but also in livestock and fisheries (Van Eenennaam 2017; Forabosco et al. 2013).

We must rethink how we manage resources, human culture and technology to create a green consumerism, with conscious consumption (Wiener and Doesher 1991). The careful choice of products and services would be crucial reducing the human impact (Moisander 2007).This can be illustrated by the UN's creation of the SDGs (General Assembly 2015), which aim to achieve large-scale sustainable and inclusive development across the globe with "nature-friendly" technologies and lifestyles (Gupta and Vegelin 2016).

### *The Anthropocentric way*

Current human behaviour and lack of investment in sustainable technologies (Gupta and Vegelin 2016) may lead to environmental collapse and crossing of planetary boundaries (Rockström et al. 2009). The main consequences are excessive damage, transformation of the Earth, increased human settlement (Boivin et al. 2016), species extinction and pollution (Costanza et al. 1997; Rhind 2009). Considering that around the pillars of the human niche is the ability to synthesise perceptual patterns through extremely complex manipulations of the environment in which we exert selective pressures (Ellis et al. 2016), often creating unsustainable conditions for the existence of other species (Surovell et al. 2005; Koch and Barnosky 2006; Johnson 2009). Furthermore, over-exploitation of forests would lead to an increase in global average temperature of almost 1°C (Feddema et al. 2005; Findell et al. 2006; Bala et al. 2007). Deforestation rates above 50% lead to ecosystem decline. In the Amazon, this leads to changes in convection, cloud formation, rainfall and solar radiation, affecting photosynthesis and plant species (Lawrence and Vandecar 2015). In the marine environment, changes in water quality, ocean acidification and changes in trophic networks due to lack of awareness and less harmful techniques are among the reasons for the decline of fish stocks (Daskalov 2002), leading to the depletion of coastal ecosystem services (Jackson et al. 2001).

The human footprint also affects us because it endangers human health through pollutants (Singh and Kumar 2024) and pesticides, combined with the consumption of contaminated food (Carbery et al. 2018). There are also restrictions on cultural and religious expressions due to the lack of local species, making it impossible to use them in rituals and cooking (Kideghesho 2009). The path of current anthropocentric 'business as usual' is certainly very dangerous for us and for the planet.

Supplementary Materials: Not applicable.

Author Contributions: Both authors equally contributed to the preparation of this paper and have read and agreed to the published version of the manuscript.

Funding: This research received no external funding

Informed Consent Statement: Not applicable.

Data Availability Statement: Not applicable.

Conflicts of Interest: The authors declare no conflict of interest.